\documentclass{revtex4}

\usepackage{graphicx}
\begin{document}
\bibliographystyle{revtex}


\title{
{\small APS/DPF/DPB Summer Study on the Future of Particle Physics,
Snowmass, CO (30 Jun--21 Jul 2001)} \\ \vspace*{12pt}
Accelerating Muons to 2400 GeV/c with Dogbones Followed by
Interleaved Fast Ramping Iron and Fixed Superconducting Magnets}



\author{D.~J.~Summers}
\email[]{summers@relativity.phy.olemiss.edu}
\thanks{Work supported by U.~S.~DOE \ DE-FG05-91ER40622}
\affiliation{University of Mississippi--Oxford, University, MS 38677 \ USA}


\date{October 15, 2001}

\begin{abstract}

The first acceleration stage for this muon collider scenario includes twenty
passes through a single two GeV Linac. Teardrop shaped arcs of 1.8 Tesla fixed
field magnets are used at each end of the Linac. This {\it dogbone} geometry
minimizes muon decay losses because muons pass through shorter arcs when their
gamma boost is low. Two 2200\,m radius hybrid rings of fixed superconducting
magnets and iron magnets ramping at 200 Hz and 330 Hz are used as part of the
second stage of muon acceleration. Muons are given 25 GeV of RF energy per
orbit.  Acceleration is from 250 GeV/c to 2400 GeV/c and requires a total of
86 orbits in both rings; 82\% of the muons survive. The total power
consumption of the iron dipoles is 4 megawatts. Stranded copper conductors and
thin magnetic laminations are used to reduce power losses.
\end{abstract}

\maketitle

\section{INTRODUCTION}

   For a $\mu^+\mu^-$ collider [1], muons must be rapidly accelerated to high
energies while minimizing the kilometers of radio frequency (RF) cavities and
magnet bores.  Cost must be moderate. Some muons may be lost to decay but not
too many. In the first stage of acceleration, consider twenty passes through
a two GeV Linac and see if enough muons survive decay.  A single continuos
Linac with teardrop shaped arcs of fixed field magnets at each end is adopted.
Muon decay losses are minimized; muons pass through shorter arcs when their
gamma boost is low.  The overall geometry looks like a dogbone [2].
More time is available for the second stage of acceleration due to the gamma 
boost.   Consider a ring of fixed superconducting magnets alternating with
iron combined function magnets rapidly cycling between full negative and full
positive field [3].  This interleaved geometry increases the average
bending field achievable in a fast ramping synchrotron and thus reduces
muon decay losses.

\section{DOGBONE LAYOUT WITH 1.8 TESLA FIXED FIELD MAGNETS}

A neutrino factory as outlined in the recent Brookhaven study [4]
provides 20 GeV muons which have enough energy to explore CP violation in the
lepton sector.  Further acceleration to 60 GeV may be enough to reach a low
mass Higgs as suggested by theory and recent measurements at LEP.

Twenty passes through a 2 GeV Linac would accelerate muons from 20 to 60 GeV.
Sets of teardrop shaped arcs as shown in Fig.~1 are used at used at 
both end of the 2 GeV linac.   
To minimize magnet cost 45$^0$ turns are used with short straight sections 
to line up the arcs.  For each teardrop, 
the length added  to the curved sections by the two straight sections is  
$(4 - 2\sqrt{2})/2\pi = 18.6\%$.
Take a muon lifetime of $2.2 \times 10^{-6}$ seconds, 1.8 Telsa dipoles,
a 70\% dipole packing fraction, and 
a 133 meter long 2 GeV Linac with 15 MV/meter.  The total magnet bore length
required is 7000 meters, 11\% longer that the Fermilab Tevatron.  
Muon
survival after twenty passes through the 2 GeV Linac 
is 95.5\%.  Squaring this percentage the luminosity is 91.8\%
of what it would be in a Higgs factory if there had been no decay loss in
accelerating the muons from 20 to 60 GeV.
The magnet cell length may have to be short to provide good acceptance for
the muons in the arcs. An alternating gradient design where the magnet
lamination change shape within a magnet avoids magnet ends and makes it
easier to consider superconducting wire rather than copper.  The magnets do
have to be at full field constantly, so power consumption is an issue.

Finally note that in a dogbone geometry, muons can orbit clockwise in one
end and counterclockwise in the other end, which may
help to preserve polarization.  If muons are 100\%  polarized, the
$\mu^+ \mu^- \to$ Higgs cross section doubles (versus the case of zero 
polarization).

\begin{figure}[t!]
\begin{center}
\vspace*{-6mm}
\resizebox{167mm}{!}{\includegraphics{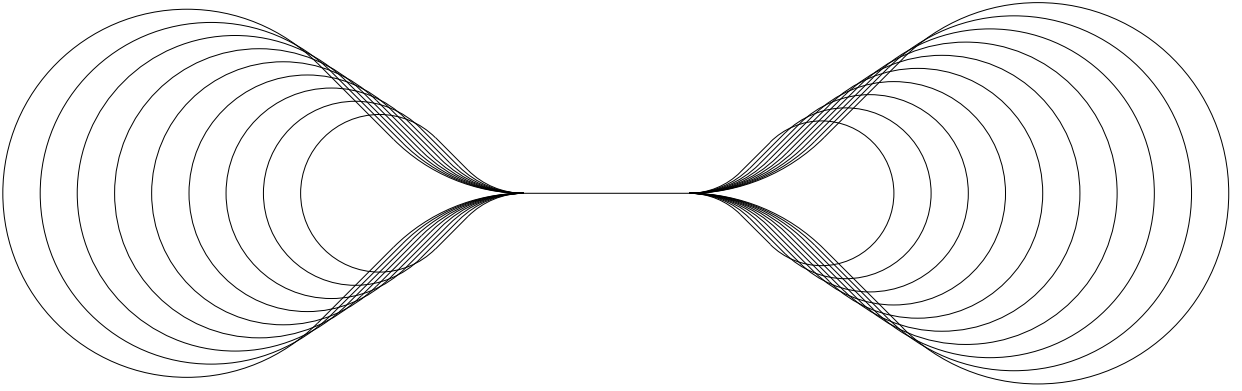}}
\vspace*{-4mm}
\caption[]{Sets of teardrop shaped magnet arcs each with 1.8 Tesla
iron dipoles are used at each end of a 2 GeV Linac.}
\label{Teardrop}
\end{center}
\end{figure}
\vspace*{-4mm}

\section{INTERLEAVED FAST RAMPING IRON AND FIXED SUPERCONDUCTING MAGNETS: 
SAGITTAS, POWER CONSUMPTION, and MUON SURVIVAL}
\vspace*{-2mm}

A lattice cell for a ring of interleaved fast ramping magnets and fixed field
superconducting magnets is shown in Fig.~2. The gradient dipoles buck the
superconducting magnets at the start of a cycle and work in unison with the
superconducting magnets at the end of a cycle. The magnetic field swings
from full negative to full positive in the gradient dipoles.

   The sagitta of a muon in a magnet increases linearly with increasing
magnetic field, $B$. It decreases linearly with increasing momentum, $p$.  And
it increases as the square of the length of a magnet, $\ell$.  The size of the
sagitta directly affects the size of magnet bores because the sagitta {\it
changes} throughout a cycle. The sagitta is given by $R - \sqrt{R^2 -
({\ell}/2)^2}$, where $R = {p / {.3 B}}$.
At 250 GeV, the sagitta is 5mm for a 2 meter long 8 Telta magnet and 11 mm
for a 6 meter long 2 Telsa magnet.
As momentum increases, the sagitta in the 8 Tesla magnets 
decreases towards
zero and the sagitta in the 2 Tesla magnets goes somewhat past zero.  

   Consider the feasibility of an iron dominated magnet which
cycles from  -2 to +2 Tesla [3].  First calculate the energy,
$W$, stored in a 2 Tesla field in a volume 6\,m long,\, .03\,m high, and 
.08\,m wide.
The permeability constant, $\mu_0$, is $4\pi\times 10^{-7}$.
$W = {B^2 / {2{\mu_0}}}[\hbox{Volume}] =$ 23\,000 Joules.
Next given 6 turns, an LC circuit capacitor, and a 250 Hz frequency; estimate 
current, voltage, inductance, and capacitance. The height, $h$, of the 
aperture is\, .03\,m.  
The top and bottom coils may be connected as two separate
circuits to halve the switching voltage.

\begin{figure}[b!]
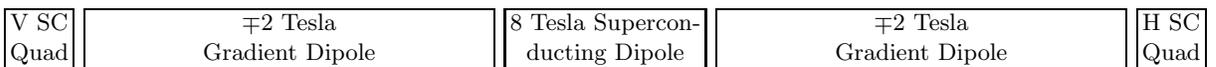

\begin{center}
\renewcommand{\arraystretch}{1.10}
\begin{tabular}{|c|c|c|c|c|c|c|c|c|}
\cline{1-1} \cline{3-3} \cline{5-5} \cline{7-7} \cline{9-9}
V SC & \quad & $\mp$2 Tesla  & \quad & 8 Tesla Supercon- & \quad &
$\mp$2 Tesla  & \quad & H SC \\
Quad      &       & \hspace*{14mm} Gradient Dipole \hspace*{14mm} 
&       & ducting Dipole &       &
\hspace*{14mm} Gradient Dipole \hspace*{14mm} &       & Quad      \\   
\cline{1-1} \cline{3-3} \cline{5-5} \cline{7-7} \cline{9-9}
\end{tabular}
\caption{Lattice cell for a ring to accelerate muons. The gradient dipole
magnetic field starts at -2 Telsa and ends at +2 Tesla. At the start of an
acceleration cycle, the gradient dipoles oppose the bending and focusing of
the superconducting dipole and quadrapoles. At the end of an acceleration
cycle, the gradient dipoles bend and focus in the unison with the
superconducting dipole and quadrapoles. $H$ signifies a horizontal quadrapole 
and $V$ signifies a vertical quadrapole.}
\label{cell}
\end{center}
\end{figure}

\begin{eqnarray}
B = {{\mu_0\,NI}\over{h}}  \quad\rightarrow\quad 
I = {{Bh}\over{\mu_0\,N}} = 8000 \ \hbox{Amps;} &
W = .5\,L\,I^2  \quad\rightarrow\quad L = {2\,W / {I^2}} = 
720\,\mu\hbox{H} \\
f = {1\over{2\pi}}\sqrt{1\over{LC}}  \quad\rightarrow\quad
C = {1\over{L\,(2\pi f)^2}} = 560\, \mu\hbox{F;} \quad & \quad
W = .5\,C\,V^2  \quad\rightarrow\quad V = \sqrt{2W / {C}} = 9000 \ 
\hbox{Volts} 
\end{eqnarray}

Now calculate the resistive energy loss, which over time is equal to one-half
the loss at the maximum current of 8000 Amps.  The one-half comes from the 
integral of cosine squared.  
Table I gives the resistivity of copper.
A six-turn copper conductor 3\,cm thick, 10\,cm high, 
and 7800\,cm long has a power dissipation of 15 kilowatts.

\begin{equation}
R = {7800 \ (1.8\,\mu\Omega\hbox{-cm})\over{(3) \, (10)}} = 470\,\mu\Omega;
\qquad
P = I^2R\int_0^{2\pi}\!\cos^2(\theta)\,d\theta = \hbox{15\,000 watts/magnet} 
\end{equation}

\begin{table}[!htb]
\begin{center}
\caption{Resistivity, magnetic saturation, and coercivity of conductors,
cooling tubes, and soft magnetic materials [9].}
\renewcommand{\arraystretch}{1.1}
\tabcolsep=3mm
\begin{tabular}{llccc} \hline \hline
Material & Composition & $\rho$ ($\mu\Omega$-cm) & B Max (Tesla) & 
H$_c$ (Oersteds)  \\
                                                                         \hline
Copper             & Cu                      & 1.8             & --- & ---   \\
Stainless 316L & Fe 70,\, Cr 18,\, Ni 10,\, Mo 2,\, C .03 & 74 & --- & ---   \\
Hastelloy B        & Ni 66,\, Mo 28,\, Fe 5  & 135             & --- & ---   \\
Titanium 6Al--4V & Ti 90,\, Al 6,\, V 4  & 171 & --- & --- \\
Titanium 8Al--1Mo--1V & Ti 90,\, Al 8,\, Mo 1,\, V 1   & 199 & --- & --- \\
Pure Iron [10]     & Fe 99.95,\, C .005      & 10              & 2.16 & .05  \\
1008 Steel         & Fe 99,\, C .08           & 12             & 2.09 &  0.8 \\
Grain--Oriented &  Si 3,\, Fe 97           & 47                & 1.95 & .1  \\
NKK Super E-Core & Si 6.5,\, Fe 93.5       & 82                & 1.8  &     \\
Metglas 2605SA1 [11] & Fe 81,\, B 14,\, Si 3,\, C 2  & 135 & 1.6    & .03  \\
Supermendur    & V 2,\, Fe 49,\, Co 49   & 26              & 2.4  & .2   \\
\hline \hline
\end{tabular}
\end{center}
\end{table}

Calculate the dissipation due to eddy currents in this conductor, which will
consist of transposed strands to reduce this loss [5--8].  
To get an idea, take the maximum B-field
during a cycle to be that generated by a 0.05m radius conductor carrying
24000 amps.  
This ignores fringe fields from the gap which will make the real answer higher.
The eddy current loss in a rectangular conductor made of transposed square 
wires 1/2 mm wide (sometimes called Litz wire) with a perpendicular magnetic
field is as follows. The width of the wire is $w$.

\begin{equation}
B = {{\mu_0\,I}\over{2\pi r}} = 0.096 \ \hbox{Tesla}; \qquad
P = \hbox{[Volume]}{{(2\pi\,f\,B\,w)^2}\over{24\rho}} 
= [.03 \ .10 \ 78]\, {{(2\pi \ 250 \ .096 \ .0005)^2} \over 
{(24)\,1.8\times{10^{-8}}}} = 3000 \ 
\hbox {watts} \\
\end{equation}

Cooling water will be needed, so calculate the eddy current losses
for cooling tubes made from type 316L stainless
steel. More exotic metals with higher resistivities are also available as shown 
in Table III. 
Choose 2 tubes per 3\,cm $\times$ 10\,cm 
stranded copper conductor for a total length of 78 $\times$ 2 = 156\,m.  
Take a 12\,mm OD
and a 10\,mm ID. Subtract the losses in the 
inner {\it missing} round conductor.
The combined eddy current loss in the copper plus the stainless steel is
4200 watts (3000 + 2400 - 1200).

\begin{eqnarray}
P(12\,\hbox{mm}) 
= \hbox{[Volume]}\, {{(2\pi\,f\,B\,d)^2} \over {32\,\rho}} 
= [\pi \ .006^2 \ 156]\,  {(2\pi\, 250 \ .096 \  .012)^2 \over 
{(32) \ 74 \times 10^{-8}}} 
= 2400 \ \hbox{watts} \\
P(10\,\hbox{mm}) 
= \hbox{[Volume]}\, {{(2\pi\,f\,B\,d)^2} \over {32\,\rho}} 
= [\pi \ .005^2 \ 156]\, {{(2\pi\,250 \ .096 \ .010)^2}
\over {(32) \ 74{\times}10^{-8}}}   
=  1200 \ \hbox{watts} 
\end{eqnarray}

   Eddy currents must be reduced in the iron not only to decrease
power consumption and cooling, but also because they introduce multipole
moments which destabilize beams.  If the laminations are longitudinal,
it is hard to force the magnetic field to be parallel to the laminations
near the gap.  This leads to additional eddy current gap losses [12].  
So consider a magnet with transverse laminations as sketched in Fig.~1
and calculate the eddy current losses. 
The yoke is either
0.28\,mm thick 3\% grain oriented silicon steel [13--15] 
or 0.025\,mm thick Metglas 2605SA1 [11].
The pole tips are 0.1\,mm thick Supermendur to raise 
the field in the gap [16].

\begin{equation}
{\hbox{P(3\% Si--Fe)} 
=  \hbox{[Volume]}{{(2\pi\,f\,B\,t)^2}\over{24\rho}}}
=  [6 \, ((.42 \ .35) - (.20 \ .23))]\, 
{{(2\pi \ 250 \ 1.6 \ .00028)^2} \over 
{(24)\,47\times{10^{-8}}}}   
=  27\,000 \ 
\hbox {watts}
\end{equation}
\begin{equation}
{\hbox{P(Metglas)}  
=  \hbox{[Volume]}{{(2\pi\,f\,B\,t)^2}\over{24\rho}}}
=  [6 \, ((.42 \ .35) - (.20 \ .23))]\, 
{{(2\pi \ 250 \ 1.6 \ .000025)^2} \over 
{(24)\,135\times{10^{-8}}}}   
=  75 \ 
\hbox {watts}  
\end{equation}
\begin{equation}
{\hbox{P(Supermendur)} 
= \hbox{[Volume]}{{(2\pi\,f\,B\,t)^2}\over{24\rho}}} 
=  [6 \, \, .09 \, \, .02]\, 
{{(2\pi \ 250 \ 2.2 \ .0001)^2} \over 
{(24)\,26\times{10^{-8}}}}   
=  210 \ 
\hbox {watts} 
\end{equation}

   Eddy currents are not the only losses in the iron.  Hysteresis 
losses,
$\int{\bf{H}}{\cdot}d\,{\bf{B}}$, scale  
with the coercive force,
H$_c$, and increase linearly with frequency.
Anomalous loss [10] which is difficult to calculate
theoretically must be included.  Thus I now use functions fitted 
to experimental measurements of 0.28\,mm thick 3\% grain oriented 
silicon steel [17], 
0.025\,mm thick Metglas 2605SA1 [11],
and 0.1\,mm thick Supermendur [17].

\begin{figure}[htb]
\begin{center}
\resizebox{167mm}{!}{\includegraphics{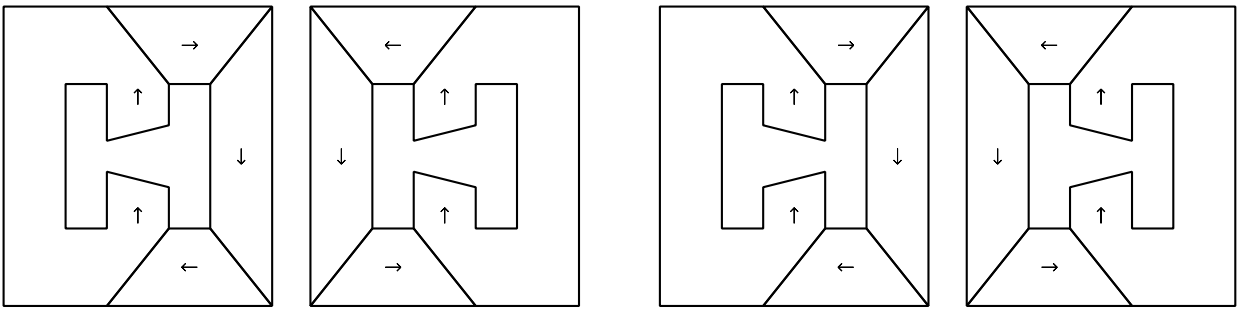}}
\caption{
Two dimensional pictures of H frame magnet laminations with grain 
oriented 3\%\,Si--Fe steel.  The 
arrows show both the magnetic field direction and the grain direction of 
the steel. Multiple pieces are used to exploit the high permeability and 
low hysteresis in the grain direction [13--15].
If Metglas 2605SA1 is used, multiple pieces are not needed.}
\end{center}
\end{figure}

\begin{equation}
\hbox{P(3\% Si--Fe)} =  4.38 \times 10^{-4} \, f^{1.67} \, B^{1.87}       
=  4.38 \times 10^{-4} \, 250^{1.67} \, 1.6^{1.87} 
=  10.7 \, \, \hbox{w/kg}                      
=  49\,000 \, \, \hbox{watts/magnet} 
\end{equation}
\vspace*{-2mm}
\begin{equation}
\hbox{P(Metglas)}    =  1.9 \times 10^{-4} \, f^{1.51} \, B^{1.74}       
=  1.9 \times 10^{-4} \, 250^{1.51} \, 1.6^{1.74} 
=  1.8 \, \, \hbox{w/kg}                         
=  7900 \, \, \hbox{watts/magnet}  
\end{equation}
\vspace*{-2mm}
\begin{equation}
\hbox{P(Supermendur)}  =  5.64 \times 10^{-3} \, f^{1.27} \, B^{1.36}       
=  5.64 \times 10^{-3} \, 250^{1.27} \, 2.2^{1.36} 
=  18 \, \, \hbox{w/kg}                         
=  1600 \, \, \hbox{watts/magnet}
\end{equation}  
\vspace*{-2mm}

\begin{table}[!htb]
\begin{minipage}{2.8 in}
\caption{Magnet core materials.}
\renewcommand{\arraystretch}{1.1}
\begin{tabular}{lcccc} \hline \hline
Material            & Thickness   & Density     & Volume     & Mass \\
                    & (mm)        & (kg/m$^3$)  & (m$^3$)    & (kg) \\ \hline
3\% Si--Fe & 0.28   & 7650        & 0.6        & 4600 \\
Metglas             & 0.025       & 7320        & 0.6        & 4400 \\
Supermendur         & 0.1         & 8150        & 0.01       & 90   \\
\hline \hline
\end{tabular}
\end{minipage}
\hspace*{15mm}
\begin{minipage}{2.9 in}
\caption{250 Hz dipole power consumption.
Eddy current components of total core losses are 27\,210 and 285
watts for 3\% Si--Fe and Metglas.}
\renewcommand{\arraystretch}{1.1}
\begin{tabular}{lcc} \hline \hline
Material                   &    3\% Si--Fe     &       Metglas    \\       
Coil Resistive Loss    & 15\,000 watts &      15\,000 watts \\
Coil Eddy Current Loss &  4200 watts      &      4200 watts  \\
Total Core Loss             &  50\,600 watts        &      9500 watts \\ \hline
Total Loss                  &  69\,800 watts &  28\,700 watts  \\ \hline \hline
\end{tabular}
\end{minipage}
\end{table}

   In summary, a 250 Hz dipole magnet close to 2 Tesla looks possible as long
as the field volume is limited and one is
willing to deal with stranded copper and thin, low hysteresis laminations.
Total losses can be held to twice the I$^2$R loss in the copper alone, 
using Metglas.

   Now with a rough design for a fast ramping magnet in hand, work out the
details of ring radii, RF requirements, and the fraction of muons
that survive decay. The fraction of the circumference packed with dipoles is
set at $P_F$ = 70\%.  As an example, consider two rings in a 2200\,m radius 
tunnel with an injection momentum of 250 GeV/c.  The first has 25\%
8T magnets and 75\% $\pm$2T magnets and ramps from 0.5T to 3.5T.
The second has 55\%
8T magnets and 45\% $\pm$2T magnets and ramps from 3.5T to 5.3T.

\begin{equation}
B = {{250\,\hbox{GeV/c}} \over {.3\,P_F\,R}} =  
{{250} \over {(.3)\,(.7)\,(2200)}} = 0.54\,\hbox{Tesla}
\end{equation}
\vspace*{-3mm}
\begin{equation}
p = (3.5\,\hbox{Tesla})\,(.3)\,(P_F)\,(R) 
  = (3.5)\,(.3)\,(.7)\,(2200) = 1600\,\hbox{GeV/c}  
\end{equation}
\begin{equation}
p =  (5.3\,\hbox{Tesla})\,(.3)\,(P_F)\,(R) 
  =  (5.3)\,(.3)\,(.7)\,(2200) = 2400\,\hbox{GeV/c} 
\end{equation}

Provide 25 GeV of RF.
The first ring accelerates muons from 250 GeV/c to 1600 GeV/c in 54 orbits.
and the second from 1600 GeV/c to 2400 GeV/c in 32 orbits.
At what frequency do the two rings have to ramp?

\begin{equation}
\hbox{Time}\,(0.5T \rightarrow 3.5T) =  {{(54)\,(2\pi)\,(2.2)} \over 
{300\,000}} 
=  2.5\,\hbox{ms}  
\qquad \rightarrow   200\, \hbox{Hz} 
\end{equation}
\vspace*{-2mm}
\begin{equation}
\hbox{Time}\,(3.5T \rightarrow 5.3T) = {{(32)\,(2\pi)\,(2.2)} \over 
{300\,000}} 
=  1.5\,\hbox{ms} 
\qquad \rightarrow 330\, \hbox{Hz} 
\end{equation}

How many muons survive during the 86 orbits from 250 GeV/c to 2400 GeV/c?
$N$ is the orbit number,
$\tau = 2.2\times10^{-6}$ is the muon lifetime, and $m = .106$ GeV/c$^2$
is the muon mass.

\begin{equation}
\hbox{SURVIVAL} = \prod_{N=1}^{86} \exp\left[{{-2\pi{R}\,m} \over 
{[250 + (25\,N)]\,c\tau}}\right] = 82\%
\end{equation}

Only 1/6 of the 18\% loss occurs in the second ring, so it is not crucial
to run it as fast as 330 Hz. 
The 250 $\rightarrow$ 1600 GeV/c ring has 1200 6\,m long dipoles
ramping at 200 Hz. The 1600 $\rightarrow$ 2400 GeV/c ring has 725 6\,m long
dipoles ramping at 330 Hz.  The weighted average rate is 250 Hz.  If
running continuously, the 1925 magnets would consume a weighted average of 29
kilowatts each for a total of 56 megawatts. But given a 15 Hz refresh rate for
the final muon storage ring [1], 
the average duty cycle for the 250 $\rightarrow$
2400 GeV/c acceleration rings is 6\%. So the power falls to 4 megawatts,
which is small.
Finally note that 
one can do a bit better than 82\% on the muon survival during final
acceleration if the first ring is
smaller, say 1000\,m, rather than 2200\,m.  

I would like to thank  S.~Berg, K.~Bourkland, S.~Bracker, R.~Fernow,
J.~Gallardo, C.~Johnstone, H.~Kirk, N.~Marks, D.~Neuffer, A.~Otter, R.~Palmer,
A.~Tollestrup, K.~Tuohy, D.~Walz, R.~Weggel, E.~Willen, and D.~Winn for their
help and suggestions.


%
%

%
%



\end{document}